\documentclass[12pt,a4paper]{article}

\usepackage{latexsym}
\usepackage{amsmath}
\usepackage{amsfonts}
\usepackage{amssymb}
\usepackage{amsthm}
\usepackage{amscd}
\usepackage{mathrsfs}
\usepackage{float}

\usepackage{curves}
\usepackage{ebezier}
\usepackage{graphics}
\usepackage{tikz}

\newcommand{\be}{\begin{equation}}
\newcommand{\ee}{\end{equation}}
\newcommand{\bea}{\begin{eqnarray}}
\newcommand{\eea}{\end{eqnarray}}

\def\fus{\circledast}   
\def\bod{{\bf D}}        %
\def\G{{\cal G}}     
\def\C{{\cal C}}                   %

\def\ifd{{      { \bf D} (G)}}         %
\def\ri{{\mathrm{i}}}                   %
\def\bR{{\mathbb R}}                    %
\def\1{{\mbox{\boldmath $1$}}}          %
\def\jp{\frac{1}{2}}                    %
\def\om{\omega}                         %

\def\noi{\noindent}         %

\begin{document}

\vspace*{0.5cm}
\begin{center}
{\Large \bf  Quasi-Hamiltonian bookkeeping of WZNW defects}
\end{center}

\vspace{0.2cm}

\begin{center}
 C. Klim\v c\'\i k   \\

\bigskip

 Institut de math\'ematiques de Luminy
 \\ 163, Avenue de Luminy \\ F-13288 Marseille, France\\
 e-mail: ctirad.klimcik@univ-amu.fr

\bigskip

\end{center}

\vspace{0.2cm}

\begin{abstract}
We interpret the  chiral WZNW model with general monodromy as an infinite dimensional quasi-Hamiltonian dynamical system.
This interpretation permits to explain  the totality of complicated cross-terms in the symplectic structures of  various  WZNW defects solely in terms of the single  concept of  the quasi-Hamiltonian fusion.  Translated from  the  WZNW language into that of the moduli space of flat connections on Riemann surfaces,  our result gives a compact and transparent  characterisation of the symplectic structure of the moduli space of flat connections on a surface
with $k$ handles, $n$ boundaries and $m$ Wilson lines.

\end{abstract}

\newpage

\section{Introduction}

The study of WZNW defects has been quite a hot topic  since last ten years \cite{AS,BG,EG,FS,FRS02,FRS03,FSW,GS,GN,GW,KO,KS,PZ,R,SZ}. The idea was  to modify the standard WZNW dynamics by consistent boundary conditions on the world-sheet or by   defect lines in the bulk  where the group valued WZNW field is allowed to jump in a particular way.
In the presence of   such defects the WZNW classical field equations can still be explicitely solved and the corresponding symplectic structure on the classical space of solutions can be   derived starting from the classical WZNW action in  \cite{G01,S}. The resulting explicit expressions for the symplectic forms  turn out to be  quite complicated,  however.

\medskip

\noi More conceptual understanding of  the WZNW symplectic structures in the presence of defects was proposed in \cite{GTT},  where the language of flat conections on Riemann surfaces was used. This insight was motivated by an older result \cite{EMSS} where  the symplectic structure of the  bulk WZNW model without defects was 
identified with that of the moduli space of flat connections on the annulus.  In the paper \cite{GTT}, the phase space of the boundary WZNW model was then  shown to be symplectomorphic to the moduli  space of flat connections on the disc with two Wilson lines inserted. The holonomies of the flat connections around the insertion points lie in some conjugacy classes in the group manifold $G$ which are  interpreted as "D-branes", i.e. as submanifolds of the target space $G$ on which the open strings   end.

\medskip

\noi Following the same philosophy, symplectic structures of several other defects were identified with those  of appropriate moduli spaces of flat connections \cite{S}. Thus the jump of the  group valued WZNW field  through a  defect line on the world-sheet  \cite{FSW,GN} was shown to lead  to  the moduli space of flat connections on the annulus with one Wilson line insertion \cite{S}.
   In this case, the holonomy around the insertion point lies in the same conjugacy class  as the jump.  The dictionnary between the WZNW
defects and the moduli spaces of flat connections was then enlarged to yet  other  types of defects still  in \cite{S}. For example, the phase space of the boundary WZNW model with one bulk defect line turns out to be the moduli space of flat connections on the disc with three Wilson line insertions. Finally, the last example treated in \cite{S} is that of permutation branes \cite{FS,FRS03,GS,R,SZ}  which are the boundary conditions for  the $n$-fold direct product $G\times G\times ... \times G$ WZNW model on a strip world-sheet. It was conjectured in \cite{S} that the relevant moduli space for this situation corresponds to the Riemann surface with $n$ boundaries and two Wilson line insertions.

\medskip

\noi Although the book-keeping of the WZNW defects via the moduli spaces of flat connections is very elegant, it is more of conceptual importance than of concrete technical utility. 
In practice,  one rather needs to have a description of the relevant symplectic structures on the moduli spaces in terms of group-valued holonomies of the flat connections since
they correspond to the physically interpretable WZNW observables. Such  description is, however, quite cumbersome already in the presence of small number of defects, since   there arise many cross terms in the symplectic forms   which correspond to   "interactions" of the defects.

\medskip

\noi The goal of the present  paper  is to propose an alternative conceptual bookkeeping of the WZNW defects which would be   technically more friendly and would use quantities with direct physical interpretation.  Our main inspiration comes from the  approach of Ref. \cite{AMM}, where the symplectic structures  of the moduli spaces of  flat connections 
on \underline{closed} surfaces (i.e. without boundaries) were described in terms of the so called quasi-Hamiltonian fusion.  Speaking more precisely, the moduli space of flat connections on the compact closed surface with  $m$ Wilson line insertions and $k$ handles was identified in \cite{AMM} 
	as the 
following symplectic manifold
\be M_{mk}\equiv  (C_1^-\fus C_2^-\fus \dots \fus C_m^-\fus \underbrace{\bod(G)\fus\cdots\fus \bod(G)}_{k \ \ {\rm times}} )_e. \label{A}\ee 
Here $C_i^-$ is the conjugacy class to which belongs  the holonomy of the connection around the $i^{th}$ insertion point (the superscript $\ ^-$ means  the inverse of the standard quasi-Hamiltonian structure on the conjugacy class), the symbols $\bod(G)$ stand for the so called internally fused quasi-Hamiltonian double of the structure Lie group $G$,   the   operation $\fus$  is the   fusion of two quasi-Hamiltonian manifolds and the notation $(M)_e$ means the \underline{symplectic} manifold obtained by the
quasi-Hamiltonian reduction of  the quasi-Hamiltonian manifold $M$ at the unit level of the moment map. 

\medskip

\noi The big advantage of the expression (\ref{A})  consists in the fact that  not only it gives the explicit characterization of the symplectic structures of the moduli spaces in terms 
of the convenient group-like variables but, at the same time, it remains conceptually neat.  Indeed, each handle or defect brings its building block into the expression and all ingredients are glued together using the single concept of the quasi-Hamiltonian fusion.

\medskip

\noi In what follows, we shall generalize the formula (\ref{A}), by allowing the presence of the boundaries on the Riemann surface. This change involves the transition from
the finite dimensional context  to an infinite-dimensional one, since the moduli spaces of flat connections  in the presence of boundaries are smooth infinite-dimensional symplectic manifolds \cite{Don}. Indeed,  the flat connections on the closed surfaces correspond  roughly to the topological $G/G$ WZNW model and the surfaces with boundaries take into account the full field theoretical WZNW dynamics.  Inspite of the   infinite-dimensional setting, the result of our generalisation is  conceptually as simple as the expression (\ref{A}). Indeed,
we shall argue that the moduli space of flat connections on the surface with  $n$  boundaries, $m$ Wilson lines insertions and $k$ handles reads:
\be M_{nmk}\equiv  (\underbrace{W^-\fus\dots \fus W^-}_{n \ \ {\rm times}}  \fus \ C^-_1\fus C^-_2\fus \dots \fus C^-_m\fus \underbrace{ \bod(G)\fus\dots\fus \bod(G)}_{k \ \ {\rm times}})_e,\label{B}\ee
where  $W^-$ is the  particular infinite-dimensional quasi-Hamiltonian manifold the points of which are quasi-periodic maps with values in $G$.  We shall refer to $W^-$  as to \underline{quasi-Hamiltonian chiral WZNW model}. 
We shall see, in particular, 
 that the quasi-Hamiltonian language of formula (\ref{B}) is very well suited   for bookkeeping of  multitude of terms appearing in the explicit description of  symplectic forms  associated to various WZNW defects.

\medskip

\noi The plan of the paper is as follows: In Section 2, we expose  some basic facts about the quasi-Hamiltonian geometry;  in particular, we define the quasi-Hamiltonian 
fusion, quasi-Hamiltonian reduction and explain the contents of the so called equivalence theorem of \cite{AMM} relating Hamiltonian loop group $LG$-spaces to  the quasi-Hamiltonian $G$-spaces.  In Section 3, we define the chiral WZNW model as the quasi-Hamiltonian system and explain how it can be obtained from the full WZNW model via that equivalence theorem just mentioned.  Section 4 prepares ingredients for proving the formula (\ref{B}), namely, it gives an elegant description of the Hamiltonian loop group space associated  by the equivalence theorem  to {\it any } quasi-Hamiltonian space.  The section 5 and 6 are respectively  devoted to the  sides $AC$  and $BC$  of the following triangle diagram  (the side $AB$ was largely discussed in \cite{S}):

\medskip

\begin{figure}[H]
  \caption{\ ~~~~~~~~~~~~~~~~~~~~~~~~~~~~~~~~~~~~~~~~~~~~~~~~~~~~~~~~~~~~~~~~~~~~~~~~~~~~~~~~~~~~~~~~~~~~~~~~~~~~~~~~~~~~~~~~~}  \label{tri}
  \begin{center}
\begin{tikzpicture}  
\draw[thick] (0,0) node[anchor= north east, rectangle, draw]{B: WZNW defects} -- (2,0) node[anchor= north west, rectangle, draw]{C: quasi-Hamiltonian geometry} 
-- (1,2) node[anchor= south, rectangle, draw]{A: Flat connections}  --cycle;
\end{tikzpicture} 
\end{center}
\end{figure}

\noi In particular, in Section 5  we  review  the  definition of the symplectic structures on the moduli space of flat connections and  then we prove that those structures are indeed described by the
formula (\ref{B}).  Finally, in Section 6, we work out the symplectic structures of the bulk, boundary and defect WZNW models starting from the formula (\ref{B}) and find agreement with the
WZNW defect symplectic structures obtained  in  \cite{G91,G01,GTT,S} by the detailed analysis of the WZNW dynamics.

\section{Quasi-Hamiltonian geometry}

Quasi-Hamiltonian manifold  $M$ is acted upon by a simple compact  connected  Lie group $G$, it is equipped with an invariant two-form $\Omega$ and with a moment map 
$\mu:M\to G$ in such a way that four  axioms must hold:

\begin{enumerate}

\item  $\mu$ intertwines the $G$ action $\triangleright$ on $M$ with the conjugacy action on $G$:

\be \mu(g\triangleright x)=g\mu(x)g^{-1}, \quad g\in G,x\in M.\label{qh1}\ee

\item The exterior derivative  of $\Omega$ is given by
\be
   \delta\Omega =-\frac{1}{12} \mu^*(\theta,[\theta,\theta]).\label{qh2}
\ee
\item The  infinitesimal action of $\G\equiv$ Lie$(G)$ on $M$  is related to $\mu$ and $\Omega$  by
\be
\iota(\zeta_M) \Omega = \jp\mu^*(\theta+\bar\theta,\zeta), \quad \forall \zeta\in\G. \label{qh3}
\ee
\item  At each $x\in M$, the kernel of $\Omega_x$ is  given by
\be
   \mathrm{Ker}(\Omega_x)=\{\zeta_M(x)\,\vert\,
   \zeta\in \mathrm{Ker}(\mathrm{Ad}_{\mu(x)}+ \operatorname{Id})\}.\label{qh4}
\ee
\end{enumerate}
Here $(.,.)$ is   the Killing-Cartan form on $\G$, 
 $\theta$ and $\bar\theta$ denote, respectively,
  the left- and right-invariant Maurer-Cartan forms on $G$ and 
  $\zeta_M$ stands for the vector field on $M$
   that corresponds to $\zeta\in\G$.   
   
\medskip

\noi Three examples of quasi-Hamiltonian manifolds will be important for us:  the conjugacy class in $G$, the  so called quasi-Hamiltonian double $D(G)$ of the group $G$ and
the internally fused double $\ifd$.
 The quasi-Hamiltonian moment map $\mu$ for  a conjugacy class  $\C\subset G$ is just the embedding $\C\hookrightarrow G$ and the quasi-Hamiltonian form $\alpha$ 
evaluated at  $f\in \C$ is defined by the formula \cite{AMM}
\be \alpha^{\C}_f(v_\xi,v_\eta)=\jp\biggl((\eta,{\rm Ad}_f\xi)-(\xi,{\rm Ad}_f\eta)\biggr).\label{fcc}\ee
Here $v_\xi, v_\eta$ are the vector fields corresponding to the infinitesimal actions of $\xi,\eta\in\G$.
There is another useful  way of representing the quasi-Hamiltonian form $\alpha$ in terms of the following parametrization of the points on the conjugacy class $\C$:
\be f=ke^{2\pi\ri\tau}k^{-1},\ee
where $\tau$ is in the Weyl alcove and $k\in G$. We have then
\be \alpha^{\C}_f=\jp(k^{-1}\delta k, e^{-2\pi\ri\tau} k^{-1}\delta k e^{2\pi\ri\tau}).\label{alt}\ee

\medskip

\noi As a manifold, the double $D(G)$ is just the direct product $G\times G$. It is the quasi-Hamiltonian $G\times G$ manifold with respect to the $G\times G$ action
\be  (g_1,g_2)\triangleright (a,b)\equiv (g_1ag_2^{-1},g_2bg_1^{-1}),\ee
moment map $\mu_D=(\mu_1,\mu_2): D(G)\to G\times G$
\be \mu_1(a,b)=ab, \quad \mu_2(a,b)=a^{-1}b^{-1}\ee
and the quasi-Hamiltonian form $\Omega_D$ defined by
\be \Omega_D=\jp(a^*\theta,b^*\bar\theta)+\jp (a^*\bar\theta,b^*\theta).\ee
As a manifold, the internally fused double $\ifd$ is  again the direct product $G\times G$ equipped with the $G$ action
\be g\triangleright (a,b)\equiv (g ag^{-1}, g bg^{-1}),\ee
the moment map
\be \mu(a,b)\equiv aba^{-1}b^{-1}\ee
and the two-form
\be \Omega =\jp (a^*\theta,b^*\bar\theta)+\jp(a^*\bar\theta,b^*\theta)+\jp((ab)^*\theta,(a^{-1}b^{-1})^*\bar\theta).\ee

\noi  Let us now list some of the   properties of the quasi-Hamiltonian spaces relevant for this paper (see \cite{AMM} for more details):
\begin{itemize}

\item  First of all,  a quasi-Hamiltonian manifold $M$ equipped with the same $G$-action, 
a  form $-\Omega$ and a moment map $\mu^{-1}$ is again quasi-Hamiltonian, it is referred to as the  inverse quasi-Hamiltonian space and denoted as $M^-$.

  \item Suppose  that the unit element $e\in G$ is the regular value of the moment map $\mu$. The  axioms of the quasi-Hamiltonian geometry
imply that $\G\equiv$ Lie($G$) acts on the unit-level submanifold
   $\mu^{-1}(e)$ without fixed points  and thus $\mu^{-1}(e)/G$ is a  symplectic  orbifold (not necessarily manifold because there still may be points in $\mu^{-1}(e)$ with  a discrete isotropy subgroup). This
   orbifold is usually denoted as $(M)_e$  and it is called the unit-level  {\it quasi-Hamiltonian reduction} of $M$. By construction,  the pull-back of the  symplectic form $\omega$
   from $(M)_e$  to $\mu^{-1}(e)$ is equal to the restriction of $\Omega$ to $\mu^{-1}(e)$, however, we stress that $\omega$ is the  symplectic form in the usual sense, whilst $\Omega$ is neither  closed nor globally non-degenerate in general.

\item A direct product of two quasi-Hamiltonian manifolds $M_1\times M_2$ is again a quasi-Hamiltonian manifold if it is equipped with the diagonal $G$-action,  a moment map
being the Lie group product $\mu_1\mu_2$ of the respective moment maps $\mu_1$  for $M_1$ and $\mu_2$ for $M_2$ and with a two-form 
\be \Omega_{12}=\Omega_1+\Omega_2+ \jp(\mu_1^*\theta,\mu_2^*\bar\theta).\label{mix}\ee
The quasi-Hamiltonian manifold $(M_1\times M_2,\mu_1\mu_2,\Omega_{12})$ is called the {\it fusion product}  and is denoted as $M_1\fus M_2$. In the case of a  multiple fusion $M_1\fus M_2\fus \dots \fus M_n$, the mixed term in (\ref{mix}) gives
rise to a multitude of terms in the resulting  reduced symplectic form  on $(M_1\fus M_2\fus \dots \fus M_n)_e$ which   look quite awkward without the conceptual quasi-Hamiltonian understanding of their origin.  It is indeed the purpose of the present paper to go in the opposite direction and to give the quasi-Hamiltonian  {\it raison d'\^etre} for the multitude of  cross-terms in the symplectic structures
induced by the WZNW defects.

\item Any $G$-invariant function $H$ on a quasi-Hamiltonian manifold $(M, \mu,\Omega)$ defines a "quasi-Hamiltonian dynamics", in the sense that there is a unique (evolution) vector
field $v_H$ satisfying the conditions
\be \iota(v_H)\Omega=\delta H, \quad \iota(v_H)\mu^*\theta=0.\ee
Here $\delta$ stands for the de Rham differential. The Hamiltonian vector field $v_H$ is $G$-invariant and preserves $\omega$ and $\mu$ \cite{AMM}. 

\item Perhaps the most remarkable property of the quasi-Hamiltonian spaces is the equivalence theorem of Ref. \cite{AMM}. It states that every quasi-Hamiltonian space
determines  a standard Hamiltonian loop group space with proper moment map and vice versa. In this way many  structural questions which can be asked about infinite-dimensional symplectic manifolds admitting the Hamiltonian actions of loop groups can  be reformulated and solved in an analytically more friendly environment, in particular, if the corresponding quasi-Hamiltonian space turns out to be finite dimensional.   Speaking  more precisely, the Hamiltonian $LG$ space $N$ with an equivariant moment map $\Phi:N\to L\G^*$
and a symplectic form $\omega$  gives rise to the quasi-Hamiltonian structure  on the manifold Hol$(N)\equiv N/\Omega G$ where $\Omega G$ is the group of based loops (i.e. loops taking
the value $e$  at the  distinguished point $\sigma=0$).
  In order to make explicite the quasi-Hamiltonian form and the  quasi-Hamiltonian moment map on Hol$(N)$, we need to introduce some technical tools, namely,  the  space of quasi-periodic maps $W$ and a map Hol$:L\G^*\to W$.

\medskip

\noi The  space $W$  consists of smooth maps   $l:\bR\to G$ with the property
\be l(\sigma+2\pi)= l(\sigma)M, \quad \forall \sigma\in\bR. \label{mon}\ee
The element $M\in G$ does not depend on $\sigma$ and it is called the monodromy of  $l\in W$.  For every $A\in L\G^*$ there si then  a unique element $w_A\in W$ such that 
\be A=w_A(\sigma)^{-1} \partial_\sigma w_A(\sigma)d\sigma, \quad w_A(0)=e.\label{ww}\ee
We have thus defined the map Hol:$L\G^*\to W$ 
\be {\rm Hol}(A):= w_A.\ee
 The loop group $LG$  acts on $L\G^*$ by gauge transformations (the Hamiltonian moment map $\Phi$ is equivariant precisely with respect to this action!):
\be g\triangleright A=gAg^{-1} -g^*\bar\theta, \quad g\in LG.\label{gt}\ee
The transformation (\ref{gt}) then induces the following transformation of the holonomy:
\be w_{g\triangleright A}(\sigma)= g(0)w_A(\sigma) g(\sigma)^{-1}.\ee
In particular, $w_A(2\pi)$ is gauge invariant with respect to the transformations from the based loop group $\Omega G$ since in this case $g(0)=g(2\pi)=e$. It is this  gauge invariance which permits to define the quasi-Hamiltonian moment
map $\mu:{\rm Hol}(N)\to G$ as 
\be \mu:=w_\Phi(2\pi).\label{mal}\ee
The quasi-Hamiltonian form $\Omega$ on Hol$(N)$ is constructed as follows.  First  of all, consider a two-form $\Upsilon$ on $L\G^*$ defined by
\be \Upsilon=\frac{1}{2}\int_0^{2\pi}d\sigma ({\rm Hol}_\sigma^*\bar\theta,\partial_\sigma{\rm Hol}_\sigma^*\bar\theta).\label{bo}\ee
 Note that the definition  (\ref{bo}) makes  sense since, for a fixed value of $\sigma$, $\textrm{Hol}_\sigma$ is a map from $L\G^*\to G$. The vector fields corresponding to the infinitesimal action of the group $\Omega G$ on $N$  turn out to be    the degeneracy directions  of the  following two-form on $N$:
 \be  \omega+\Phi^*\Upsilon.\label{qhf}\ee
 This two-form is therefore the pull-back of some form $\Omega$ on Hol$(N)$, which is nothing but the quasi-Hamiltonian  form on Hol$(N)$.

\ 
\end{itemize}

\section{Quasi-Hamiltonian equivalent of the WZNW model}

The full WZNW model \cite{W} is the standard symplectic dynamical system, the phase space $P_{WZ}$ of which admits two different Hamiltonian actions of the loop group $LG$.
One of those actions has the  equivariant moment map in the sense of Definition 8.2 of \cite{AMM}  (see also Eq. (\ref{emm}) of the present paper). Following the discussion at the end of  Section 2, we can associate to the equivariant Hamiltonian $LG$-manifold $P_{WZ}$  the equivalent quasi-Hamiltonian dynamical system on the space Hol$(P_{WZ})$ equipped with  the corresponding $G$-action induced by some quasi-Hamiltonian moment map. It is the goal of this section to show that this equivalent quasi-Hamiltonian system  is nothing but the   quasi-Hamiltonian version of the chiral WZNW model.

\medskip

\noi Ideologically, we shall describe here the WZNW model in the language of the twisted Heisenberg double \cite{ST,K}.  Thus the phase space $P_{WZ}$ of the WZNW model is the cotangent bundle of the loop group $LG$ parametrized by $J_L(\sigma)\in L\G$ and $g(\sigma)\in LG$, however, the symplectic form 
is not the canonical one  on the cotangent bundle since it contains  the additional term (the twist):
\be \omega_{WZ}= -\delta\int_0^{2\pi} d\sigma(J_L,\delta gg^{-1})-\frac{1}{2}\int_0^{2\pi} d\sigma(\delta gg^{-1},\partial_\sigma (\delta gg^{-1})).\ee
Here 
$\delta$ is the de Rham differential on $P_{WZ}$.

\medskip

\noi There are two Hamiltonian actions of the loop group $LG$ on the phase space $P_{WZ}$:
\be\begin{aligned} h\triangleright _L(J_L,g)&:= (hJ_Lh^{-1}+\partial_\sigma hh^{-1}, hg), \qquad h\in LG;\\
 h\triangleright_R (J_L,g)&:=(J_L, gh^{-1}), \quad ~~~~~~~~~\ \ ~~~~~~~~~ h\in LG.\end{aligned}\label{raa}\ee
 The moment maps of these two actions are $J_L$ and $J_R$, respectively, where
 \be J_R:= -g^{-1}J_Lg+g^{-1}\partial_\sigma g.\label{rl}\ee
 Indeed, it is  easy to check that it holds
 \be \iota(v^L_\xi)\omega_{WZ}=\delta\int_0^{2\pi}(J_L,\xi)d\sigma,\qquad  \iota(v^R_\xi)\omega_{WZ}=\delta\int_0^{2\pi}(J_R,\xi)d\sigma,\ee
where $\xi\in L\G$ and $v^{L,R}_\xi$ are the respective vector fields corresponding to the infinitesimal actions of $\xi$ on $P_{WZ}$.
Note the transformation of the right current $J_R$ under the action $\triangleright_R$ on $P_{WZ}$  by an element $h\in LG$:
\be J_R\to hJ_Rh^{-1}-\partial hh^{-1}.\label{emm}\ee
We observe that the moment map $J_R$ is equivariant following the conventions of Sections 8.1 and 8.2 of \cite{AMM}. However, the left current $J_L$ transforms under the
action $\triangleright_L$ with the opposite sign of the inhomogeneous term:
\be J_L\to hJ_Lh^{-1}+\partial hh^{-1}.\label{aemm}\ee
We shall refer to the moment map $J_L$ as 'anti-equivariant'.  We finish the resuming of the WZNW model by defining its Hamiltonian:
\be H_{WZ}=-\jp\int_0^{2\pi}(J_L,J_L) d\sigma- \jp\int_0^{2\pi}(J_R,J_R) d\sigma.\ee

\medskip

\noi Since the  phase space $P_{WZ}$  with the right action $\triangleright_R$ of $LG$ is the Hamiltonian $LG$-space  in the sense of the definition 8.2 of \cite{AMM}, we can construct
the corresponding quasi-Hamiltonian $G$-space  Hol$(P_{WZ})$ following the recipe described at the end of Section 2.  This gives the statement of the following important
Theorem:

\medskip

\noi {\bf Theorem 1}: {\it The quasi-Hamiltonian space {\rm Hol}$(P_{WZ})$ is the space of quasi-periodic maps $W$, the corresponding quasi-Hamiltonian moment map $\mu:W\to G$ is the inverse  monodromy of the
element $l\in W$
\be \mu(l)=l(2\pi)^{-1}l(0)\label{lmn}\ee
and the quasi-Hamiltonian form $\Omega$ on $W$ induced by $\omega_{WZ}$ on $P_{WZ}$ reads
 \be \Omega(l):=\frac{1}{ 2}\biggl[\int_0^{2\pi}(l^{-1}\delta l, \partial_\sigma(l^{-1}\delta l))d\sigma +(\delta l l^{-1}\vert_0, \delta l l^{-1}\vert_{2\pi})\biggr].\label{foo}\ee}
 \begin{proof} Denote by $g_R\in W$ the element $w_{J_R}$ defined by (\ref{ww}), i.e.
 \be J_R= g_R^{-1}\partial_\sigma g_R, \qquad g_R(0)=e.\ee
 We can also conveniently parametrize the current $J_L$ as
 \be J_L=\partial_\sigma g_Lg_L^{-1}, \qquad g_L(0)=e\ee
 and the field $g(\sigma)$ as 
 \be g(\sigma)=g_L(\sigma)b(\sigma)g_R(\sigma).\label{pab}\ee
 The relation (\ref{rl}) then implies that $b(\sigma)$ in fact does not depend on $\sigma$ and it is therefore  equal to $g(0)$. In what follows, we set 
 \be l(\sigma):=g_L(\sigma)b(\sigma)=g_L(\sigma)g(0)\label{jr}\ee
 and express straightforwardly   the symplectic form $\omega_{WZ}$ in terms of the variables $l(\sigma)$ and $g_R(\sigma)$:
\be \omega_{WZ} = \frac{1}{2}\int_0^{2\pi}(l^{-1}\delta l,\partial_\sigma(l^{-1}\delta l))-    \frac{1}{2} (l^{-1}\delta l,\delta g_R g_R^{-1})\bigg\vert_0^{2\pi}-\frac{1}{2} 
\int_0^{2\pi}(\delta g_R g_R^{-1},\partial_\sigma (\delta g_R g_R^{-1})).               \ee
Because of the fact that $g_R(0)=e$ and \be g(2\pi)= l(2\pi)g_R(2\pi)=l(0)=g(0),\label{esm}\ee we conclude that the quasi-Hamiltonian form (\ref{qhf}) becomes
\be \omega_{WZ}+J_R^*\Upsilon= \omega_{WZ}+\frac{1}{2} 
\int_0^{2\pi}(\delta g_R g_R^{-1},\partial_\sigma (\delta g_R g_R^{-1}))= \frac{1}{ 2}\biggl[\int_0^{2\pi}(l^{-1}\delta l, \partial_\sigma(l^{-1}\delta l))d\sigma +(\delta l l^{-1}\vert_0, \delta l l^{-1}\vert_{2\pi})\biggr].         \ee
Now Eqs. (\ref{mal}) and  (\ref{esm}) show that the quasi-Hamiltonian moment map is indeed the inverse monodromy of $l\in W$
\be \mu(l)=g_R(2\pi)=l(2\pi)^{-1}l(0).\label{lmm}\ee
Finally,  it remains  to identify the quasi-Hamiltonian space Hol$(P_{WZ})$ with $W$. Note that Hol$(P_{WZ})$ is the space of cosets $P_{WZ}/\Omega G$, so starting from the parametrization $(J_L,g)$ of $P_{WZ}$ we see that Hol$(P_{WZ})$  can be parametrized by means of $g_L$ and $g(0)$  as $J_L=\partial_\sigma g_Lg_L^{-1}$, $g=g(0)$.
Following (\ref{jr}), Hol$(P_{WZ})$ can be parametrized also by $l\in W$ since $g(0)=l(0)$. From (\ref{raa}), we conclude that the $G$-action on $W$ is given by
\be l(\sigma)\to l(\sigma)h^{-1}, \qquad l(\sigma)\in W, \quad h\in G.\label{uzr}\ee
 
 \end{proof}

\noi Although from  the general theorems of  Ref. \cite{AMM} it follows that  the triple $(W,\Omega(l),\mu(l))$ given by Eqs. (\ref{foo}), (\ref{lmm}) and (\ref{uzr}) is the quasi-Hamiltonian 
$G$-space, we prefer to provide a direct proof of this fact in order  to  make the present  paper more self-contained:

\medskip

   \noi {\bf Theorem 2}:  {\it  Define a function on $W$ by the formula 
   \be H(l)=-\jp\int_0^{2\pi}(\partial_\sigma ll^{-1},\partial_\sigma ll^{-1})d\sigma,\label{ham}\ee
   the $G$-action  on $W$ by 
   \be l(\sigma)\to l(\sigma)h^{-1}, \qquad l(\sigma)\in W, \quad h\in G,\label{uzr2}\ee
   the moment map  $\mu:W\to G$ by 
   \be \mu(l)=l(2\pi)^{-1}l(0)\label{lmm2}\ee
   and the two-form $\Omega(l)$ on $W$ by 
   \be \Omega(l):=\frac{1}{ 2}\int_0^{2\pi}(l^{-1}\delta l, \partial_\sigma(l^{-1}\delta l))d\sigma +\jp(\delta l l^{-1}\vert_0, \delta l l^{-1}\vert_{2\pi})\label{foo2}\ee
   The  quadruple  $(W,\Omega(l),\mu,H) $ is  then the  quasi-Hamiltonian dynamical system.}

 \begin{proof}
We immediately observe from (\ref{mon}), (\ref{uzr2})  and  (\ref{lmm2}) that
\be \mu(h\triangleright l)=h\mu(l)h^{-1},\ee
which means that the  first defining quasi-Hamiltonian property (\ref{qh1}) is verified.

\medskip

\noi  A simple bookkeeping of boundary terms gives  the  second defining quasi-Hamiltonian property (\ref{qh2}):
$$ \delta\Omega(l)= \frac{1}{12} (l^{-1}\delta l\vert_{2\pi},[l^{-1}\delta l\vert_{2\pi},l^{-1}\delta l\vert_{2\pi}])- \frac{1}{12} (l^{-1}\delta l\vert_{0},[l^{-1}\delta l\vert_{0},l^{-1}\delta l\vert_{0}])+\jp \delta (\delta l l^{-1}\vert_0, \delta l l^{-1}\vert_{2\pi})=$$
\be    =-\frac{1}{12} (\mu_l^{-1}\delta\mu_l,  [\mu_l^{-1}\delta\mu_l,\mu_l^{-1}\delta\mu_l]),  \label{mmm}\ee
where we have set $\mu(l)=\mu_l$.

\medskip

\noi   Let us now verify the third property (\ref{qh3}). First of all, let $\xi_{W}$ be a vector field on $W$ induced by the infinitesimal action of an  element $\xi\in\G$.
We infer easily 
\be \iota(\xi_{W})l^{-1}\delta l=-\xi, \quad \iota(\xi_{W})(\delta l l^{-1}\vert_{0})=-l(0)\xi l(0)^{-1}, \quad \iota(\xi _{W})(\delta l l^{-1}\vert_{2\pi})=-l(2\pi)\xi l(2\pi)^{-1},\ee
hence we find indeed that 
\be \iota(\xi_W)\Omega=\jp (\xi ,\delta \mu_l\mu_l^{-1}+\mu_l^{-1}\delta \mu_l).\ee

\medskip

\noi It remains to verify the last property (\ref{qh4}). First of all we note that $W$ is a submanifold of the group $\bR G$ consisting  of all smooth maps from $\bR$ to $G$.
Therefore any vector field $v$ at a point $l$ of $W$ can be written as the left transport  $-L_{l*}\zeta$ of some $-\zeta\in$ Lie$(\bR G)$. From this information we find
\be \iota(v)(l^{-1}\delta l)=-\zeta\ee
therefore
\be \iota(v)\Omega= \int_0^{2\pi}(l^{-1}\delta l,\partial_{\sigma}\zeta) d\sigma -\jp(\zeta,l^{-1}\delta l)\bigg\vert^{2\pi}_0 -\jp(l_0\zeta_0l_0^{-1},\delta ll^{-1}\vert_{2\pi}) +\jp
(\delta ll^{-1}\vert_0, l_{2\pi}\zeta_{2\pi}l_{2\pi}^{-1}).\ee
If $v$ is to be in the kernel of $\Omega$ then obviously $\partial_\sigma\zeta=0$ and
\be \iota(v)\Omega=\jp (\zeta ,\delta\mu_l\mu_l^{-1}+\mu_l^{-1}\delta\mu_l) = \jp( \mu_l\zeta \mu_l^{-1} +\zeta, \delta\mu_l\mu_l^{-1}).\ee
From the last formula, the wanted property (\ref{qh4}) readily follows.

\medskip

\noi We conclude the demonstration by noting that the Hamiltonian (\ref{ham}) is evidently $G$-invariant, as it should be.

\end{proof}

\noi {\bf Definition}: {\it  We shall refer to the  quasi-Hamiltonian dynamical system  $(W,\Omega(l),\mu(l),H(l)) $ as to the quasi-Hamiltonian chiral WZNW model.}

\medskip

\noi {\bf Remark 1}:  {\footnotesize Historically, the origin of the concept of the chiral WZNW model  lies in the attempt to equip   the left and right movers of the WZNW model
with independent  dynamics.  Recall that every solution of the WZNW model in the configuration space $LG$ can be described as the  product of left and right movers \cite{G91,G01}:
\be g(\sigma,\tau)=l(\sigma+\tau)r^{-1}(\sigma-\tau),  \qquad \sigma\in[0,2\pi[, \quad \tau\in\bR,\ee
where both left and right movers $l$ and $r$ are  the elements of $W$   and can be viewed as  almost  independent coordinates on  the infinite-dimensional  phase space $P_{WZ}$  of the theory. Indeed, $l$ and $r$  are tied only by the requirement that they must have {\it the same  monodromies} in order to insure the periodicity of the  WZNW   field $g(\sigma)$. 
In \cite{G91},   the symplectic form $\omega_{WZ}$  was expressed in terms of the left and right movers as 
\be \omega_{WZ}=\Omega(l) -\Omega(r),\label{gaw}\ee
where the two-form $\Omega(l)$ is nothing but our quasi-Hamiltonian friend (\ref{foo}).  The form of the WZNW symplectic form (\ref{gaw}) suggests that it may be possible  
 to separate completely the left and right movers by allowing the independent monodromies for them.  However, the trouble in doing that was remarked already in \cite{G91}. The point is that the  exterior derivatives of 
the
forms $\Omega(l)$ and $\Omega(r)$ do not vanish separately   as it can be seen from (\ref{mmm}) (in fact, in calculating $\delta\omega_{WZ}$, they cancel with each  other  precisely   when the left and right monodromies are the same).
As the  solution to the problem of non-closedness of $\Omega(l)$, it was proposed in \cite{G91}   to add to $\Omega(l)$ a two-form $\rho(\mu(l))$ depending exclusively on the 
inverse monodromy $\mu(l)$ and  to define the chiral WZNW model as a theory on the phase space $W$, with the symplectic form $\Omega(l)+\rho(\mu(l))$ and the  quadratic current Hamiltonian (\ref{ham}). The problem with this definition is the ambiguity of the choice of the two-form $\rho(\mu(l))$ as well as the fact that, strictly speaking, such $\rho$ exists only on a dense open subset
of the group manifold $G$. 
In this section, we  did not attempt to define the   chiral dynamics in the  symplectic way, but we  adopted  the quasi-Hamiltonian point of view. Said in other words, we  have defined  the chiral  WZNW model as the {\it quasi-Hamiltonian dynamical system}. For this, we did not need to add  any term to the  two-form $\Omega(l)$ on $W$,  but we let it as it stands. Of course, all this is just a shift of interpretation but  it will turn out soon that our quasi-Hamiltonian version of the chiral WZNW has some good structural  properties, namely it is useful  for the compact  description of the symplectic properties of the WZNW defects.   }

 \section{Loop group equivalent of a  quasi-Hamiltonian space}
 
 \noi We devote this section  to the  formulation and proof  of a  technical  Theorem 3, which will be of big utility in Section 5.   It gives a convenient description of the Hamiltonian $LG$-space equivalent to a given quasi-Hamiltonian space in the sense of the equivalence formulated in Section 2:
 
 \noi {\bf Theorem 3}: {\it The Hamiltonian  $LG$-manifold $(N,\omega,\Phi)$ with equivariant proper moment map equivalent to a quasi-Hamiltonian $G$-manifold $(M,\Omega,\mu)$   is given by $N=(M\fus W^-)_e$, i.e. by the quasi-Hamiltonian fusion of $M$ and $W^-$ 
 followed by the unit-level quasi-Hamiltonian reduction. The corresponding $LG$-action on $(M\fus W^-)_e$ is given by
 \be (x,l(\sigma))\to (x,h(\sigma)l(\sigma)), \qquad x\in M, \quad l(\sigma)\in W, \quad h(\sigma)\in LG\label{lga}\ee
 and the corresponding  $L\G^*$-valued moment map  $\Phi$ is given by 
 \be \Phi(x,l)=-\partial_\sigma ll^{-1}d\sigma, \qquad x\in M,\quad l\in W.\label{3mm}\ee}
 
 \begin{proof}
We start by checking,  that  the formula (\ref{lga}) consistently defines the $LG$-action on the quasi-Hamiltonian quotient $(M\fus W^-)_e$. First of all,  the monodromy 
 of the configuration $h(\sigma)l(\sigma)$ is the same as that of $l(\sigma)$ for every $h(\sigma)\in LG$ therefore the action (\ref{lga}) survives the unit-level reduction constraint 
 $\mu\mu_l^{-1}=e$. On the top of that, the action (\ref{lga}) obviously commutes with the quasi-Hamiltonian $G$-action (\ref{uzr2}) on $W$, it  descends therefore to the
 $G$-quotient.   
 
 \medskip
 
 \noi In what follows, we find more convenient to describe the space $(M\fus W^-)_e$ differently.    For that, consider the quasi-Hamiltonian $G\times G$ action
 on $M\times W^-$, that is the $G$-action on $M$ and the action (\ref{uzr2}) on $W^-$.  Now the diagonal subaction, the quotient with respect to which we consider, permits
 a global slice given by the requirement $l(0)=e$.  We shall denote by $\tilde l$ the elements of $W$ for which this requirement is respected, i.e.  $\tilde l(0)=0$ and
 we parametrize $(M\fus W^-)_e$  as
 \be  (M\fus W^-)_e =\{(x,\tilde l)\in M\times W, \quad \tilde l(0)=e,\quad \mu(x)\tilde l(2\pi)=e\}. \label{spa}\ee
We infer from (\ref{foo}) that,  in the parametrization (\ref{spa}), the symplectic form $\omega$ on $(M\fus W^-)_e$ obtained form the quasi-Hamiltonian reduction reads
 \be \omega =\Omega- \frac{1}{ 2}\int_0^{2\pi}(\tilde l^{-1}\delta \tilde l, \partial_\sigma(\tilde l^{-1}\delta \tilde l))d\sigma.\ee
 In order to verify that (\ref{3mm}) gives the moment map of the $LG$-action (\ref{lga}), we have to characterize this action in the parametrization (\ref{spa}). We distinguish
 two cases: the action of the based loops from $\Omega G$ and the action of the constant loops from $G$.  We find
 \be (x,\tilde l)\to (x,h\tilde l), \quad  h\in\Omega G;\label{aca}\ee
 \be (x,\tilde l)\to (h\triangleright x,  h\tilde l h^{-1}), \quad h\in G. \label{acb}\ee
  Here $h\triangleright x$ stands for the $G$-action on $M$.
  
  \medskip
  
  \noi Denote by $v_\xi$ the vector field corresponding to the infinitesimal action (\ref{aca}) of an element $\xi\in$ Lie$(\Omega G)$. Then we find easily
  $$ \iota(v_\xi)\omega=-\jp\int_0^{2\pi}(\tilde l^{-1}\xi\tilde l, \partial_\sigma(\tilde l^{-1}\delta \tilde l))d\sigma +\jp\int_0^{2\pi}(\tilde l^{-1}\delta \tilde l, \partial_\sigma(\tilde l^{-1}\xi\tilde l))d\sigma =$$\be = -\int_0^{2\pi}(\xi, \tilde l\partial_\sigma(\tilde l^{-1}\delta \tilde l)\tilde l^{-1})d\sigma= -\delta \int_0^{2\pi}(\xi,  \partial_\sigma\tilde l \tilde l^{-1})d\sigma\label{prr}\ee
  We note that all boundary terms in the computation (\ref{prr}) vanished because of $\xi(0)=0$.
  
  \medskip
  
  \noi Denote by $v_\xi$ the vector field corresponding to the infinitesimal action (\ref{acb}) of an element $\xi\in$Lie$(G)$. Then we  find easily
  $$ \iota(v_\xi)\omega=\iota(v_\xi)\Omega-\jp\int_0^{2\pi}(\tilde l^{-1}\xi\tilde l-\xi,\partial_\sigma(\tilde l^{-1}\delta \tilde l))+\jp \int_0^{2\pi} (\tilde l^{-1}\delta \tilde l, \partial_\sigma
  (\tilde l^{-1}\xi\tilde l-\xi))=$$ $$ = \jp(\xi,\delta\mu\mu^{-1}+\mu^{-1}\delta\mu)+\jp(\xi,\tilde l^{-1}\delta \tilde l)\vert_0^{2\pi}+\jp(\tilde l^{-1}\delta \tilde l,\tilde l^{-1}\xi\tilde l)\vert_0^{2\pi}-\int_0^{2\pi}(\xi, \tilde l\partial_\sigma( \tilde l^{-1}\delta \tilde l)\tilde l^{-1})=$$ \be = \jp(\xi,\delta\mu\mu^{-1}+\mu^{-1}\delta\mu)+\jp(\xi,\delta\tilde l(2\pi)\tilde l(2\pi)^{-1}+\tilde l(2\pi)^{-1}\delta\tilde l(2\pi))-\delta \int_0^{2\pi}(\xi,  \partial_\sigma\tilde l \tilde l^{-1})d\sigma.\label{46x}\ee
 Following (\ref{spa}),  the first two terms on the r.h.s. of (\ref{46x}) vanish because of the constraint  $ \mu(x)\tilde l(2\pi)=e$.  Combining this fact with  (\ref{prr}), we conclude that
 $\Phi=-\partial_\sigma \tilde l\tilde l^{-1}d\sigma\in L\G^*$ is indeed the moment map of the $LG$-action (\ref{lga}) on the symplectic manifold $(M\fus W^-)_e$.
 
 \medskip
 
 \noi It remains to prove that the  Hamiltonian  $LG$-space $((M\fus W^-)_e,\omega,\Phi)$ is equivalent to the quasi-Hamiltonian $G$-space $(M,\Omega,\mu)$ in the sense of the equivalence
  discussed
 at the end of Section 2. For that we shall determine the equivalent system $(M',\Omega',\mu')$ to $((M\fus W^-)_e,\omega,\Phi)$ and then show that $(M',\Omega',\mu')$ and $(M,\Omega,\mu)$
 are isomorphic as the quasi-Hamiltonian spaces. Let us first prove that the quotient  $M'\equiv (M\fus W^-)_e/\Omega G$ indeed coincides with $M$ as manifold. For that, we use the following well-known parametrization  of the space $W$ of quasi-periodic maps used in \cite{G91,G01}:
 \be l(\sigma)\equiv h(\sigma)e^{\ri \tau \sigma} g_0^{-1},\label{bla}\ee
 where $h(\sigma)\in LG$, $g_0\in G$ and $\tau$ is the element of the Weyl alcove.
 It follows from (\ref{bla}), in  particular, that  the elements $\tilde l(\sigma)$ can be parametrized as
  \be l(\sigma)\equiv k(\sigma)g_0e^{\ri \tau \sigma} g_0^{-1},\label{blaa}\ee
  where $k(\sigma)$ is in the based loop group $\Omega G$. The quotient $M'=(M\fus W^-)_e/\Omega G$ can be therefore identified  with the set of elements $(x,g_0e^{\ri \tau \sigma} g_0^{-1})\in M\times W$   such that $g_0e^{\ri 2\pi \tau } g_0^{-1}=\mu(x)^{-1}$ and this set, in turn, can be directly identified with $M$.
  
  \medskip
  
  \noi Following Eq. (\ref{qhf}), the quasi-Hamiltonian form $\Omega'$ on $M$ which corresponds to the Hamiltonian  $LG$-space $((M\fus W^-)_e,\omega,\Phi)$
  is given by the formula
 \be \Omega'=\omega-\frac{1}{2} \int_0^{2\pi}(\delta g_R g_R^{-1},\partial_\sigma (\delta g_R g_R^{-1}))=\Omega- \frac{1}{ 2}\int_0^{2\pi}(\tilde l^{-1}\delta \tilde l,  \partial_\sigma(\tilde l^{-1}\delta \tilde l))d\sigma-\frac{1}{2} \int_0^{2\pi}(\delta g_R g_R^{-1},\partial_\sigma (\delta g_R g_R^{-1})),\ee
where  $g_R\in W$, $g_R(0)=e$ is defined by
\be g_R^{-1}\partial_\sigma g_R d\sigma=\Phi= -\partial_\sigma\tilde l\tilde l^{-1}d\sigma.\ee
  We thus see that $g_R=\tilde l^{-1}$ and
  \be \Omega'=\Omega.\ee
  The fact that the induced $G$-action on $M'$ is clearly given  by the restriction of the action (\ref{acb}) on the first term, i.e. by the $G$-action on $M$. Finally, the moment map $\mu'$
  is given by $g_R(2\pi)=\tilde l(2\pi)^{-1}=\mu(x)$ which finishes the proof.
 
 \end{proof}
 
 \noi Repeating step by step the proof of Theorem 3, we obtain also
 
 \medskip 
 
 \noi{\bf Corollary}: {\it  The manifold $(M\fus W)_e$ is the Hamiltonian $LG$-space with the $LG$-action given by (\ref{lga}) and the anti-equivariant  moment map given  by \be   \Phi^- =\partial_\sigma ll^{-1}d\sigma.\ee}

\medskip

\noi {\bf Remark 2}:   Theorem 3 can be easily generalized to the case where the manifold $M$ is the quasi-Hamiltonian $G\times G$-space and we transform to the loop group language only 
one copy of $G$.  We find then that the resulting fusion/reduction $(M\fus W^-)_e$ does not give a symplectic manifold but it yields the quasi-Hamiltonian $G$-space with respect to the copy of $G$ which we "did not touch".  For example, if $M$ is the quasi-Hamiltonian double $D(G)$ then a one-line   computation shows  that $(D(G)\fus W^-)_e=W^-$, or, said in other words, $D(G)$ acts as identity with respect to the partial fusion.

\medskip

\noi   To a given  Hamiltonian $LG$-space $(N,\omega)$ with the equivariant moment map $\Phi$, one can canonically associate its "loop-reversal"   Hamiltonian  $LG$-space $(N,\omega)$
with the anti-equivariant moment map $\Phi^-$ (the term anti-equivariant was defined by means of  Eq. (\ref{aemm})).   To do that, we define  the loop-reversal map $I: S^1\to S^1$ as 
\be I(\sigma)= 2\pi-\sigma, \qquad \sigma\in[0,2\pi[. \label{lrm}\ee
Let now act the loop group $LG$ on $N$ as
\be h\triangleright_I y:=(I^*h)\triangleright y, \qquad h\in LG, \quad y\in N,\label{ia}\ee
where $\triangleright$ stands for the original $LG$-action with the equivariant moment map $\Phi$ and $\triangleright_I$ stands for the new action defined in terms of the original one and of the pull-back $I^*$ of the map $I$.  It is easy to see that the new action  $\triangleright_I$ has  the  anti-equivariant moment map $ \Phi^-=-I^*\Phi$.
Indeed, we have
\be \iota(I^*\xi)\omega=\delta\int (\Phi,I^*\xi) =\delta\int(-I^*\Phi,\xi).\ee
We have  now the following proposition

\medskip

\noi {\bf Theorem 4}: {\it  The anti-equivariant Hamiltonian $LG$-space $(M\fus W)_e$ is isomorphic to  the loop reversal of the equivariant space $(M\fus W^-)_e$.}

 \begin{proof} The quasi-Hamiltonian space $(W,\Omega(l),\mu(l))$ corresponding to the chiral WZNW model has an interesting property that its quasi-Hamiltonian inverse $(W,-\Omega(l),\mu(l)^{-1})$ is isomorphic to the
original space $(W,\Omega(l),\mu(l))$. This  isomorphism $I^*:W\to W$ is simply  the extension of the pull-back of the  loop reversal map and, with a slight abuse of notation, we have denoted it
again by $I^*$:
 \be (I^*l)(\sigma):= l(2\pi-\sigma).\ee
To see that this is isomorphism, we just check    that the $G$-action (\ref{uzr2}) commutes with $I^*$ and it holds $I^*\Omega=-\Omega$ and $\mu(I(l))=\mu(l)^{-1}$. 
  
  \medskip
  
  \noi The existence of the isomorphism $I^*$  obviously implies   that 
 the reduction/fusion $(M\fus W)_e$ is isomorphic to 
$(M\fus W^-)_e$ as symplectic manifold, but not necessarily as $LG$-space.  
 Indeed,  the symplectic form $\Omega+\Omega(l)$ on  $(M\fus W)_e$  can be rewritten in the coordinates
$\hat l\equiv I^*l$ on $W$ as $\Omega-\Omega(\hat l)$  and in the same coordinates  the action (\ref{lga}) of the loop group becomes the action
 \be (x,\hat l(\sigma))\to (x,h(2\pi-\sigma)\hat l(\sigma)), \qquad x\in M, \quad \hat l(\sigma)\in W, \quad h(\sigma)\in LG\label{lgaa}.\ee
 Thus we see that the change of coordinates $l\to \hat l$ gives  the loop reversed action $\triangleright_I$ of $LG$ on $(M\fus W^-)_e$.  
  \end{proof}

\section{Flat connections and the proof of  formula (\ref{B})}

In this section, we wish to deal with the side $AC$ of the triangle on Fig. 1 and to  prove the formula (\ref{B}).

\medskip

\noi Let $G$ be a compact simple connected and simply connected Lie group, $\G$ its Lie algebra  and $\Sigma $ be a Riemann surface with boundaries $\partial \Sigma$. Denote by $G(\Sigma)$ the group of smooth maps from  $\Sigma$ to $G$. The group $G(\Sigma)$ naturally acts  on the space of connections on the trivial bundle $\Sigma\times G$ which we denote as $\Omega^1(\Sigma,\G)$:
\be A^g=gAg^{-1}-g^*(\bar \theta),\qquad A\in \Omega^1(\Sigma,\G), \quad  g\in G(\Sigma). \label{cac} \ee
 The space  $\Omega^1(\Sigma,\G)$ is symplectic; its symplectic form $\omega$  is  defined by  
 \be \omega=\int_\Sigma(\delta A\stackrel{\wedge}{,}\delta A),\ee
 where $\delta$ stands for the de Rham differential on the infinite-dimensional manifold  $\Omega^1(\Sigma,\G)$ and $(.,.)$ is the Killing-Cartan form on $\G$. It turns out \cite{AB}  that
 the action (\ref{cac}) is symplectic  with the moment map  $\Psi$ given by 
 \be \langle \Psi(A),\xi\rangle\equiv \int_\Sigma(dA+A^2,\xi) +\int_{\partial\Sigma}(A,\xi),\label{mmp}\ee
 where $\xi\in$Lie$(G(\Sigma))\equiv  \Omega^0(\Sigma,\G)$ and $d$ is the de Rham differential on the surface $\Sigma$.  Note that we take the orientation
 on $\partial\Sigma$ opposite to the induced orientation on $\Sigma$ as in \cite{AMM,MW2}. 
 
 \medskip
 
 \noi The object of central interest for us is obtained by a partial symplectic reduction of the full connection space $\Omega^1(\Sigma,\G)$ by the subgroup
   $G_\partial (\Sigma)$   of $G(\Sigma)$ consisting of map sending the boundaries to the unit element $e$ of $G$. The moment map of this action  is given just by the first
  term in (\ref{mmp})  with $\xi \in  $ Lie$(G_\partial(\Sigma))\subset \Omega^0(\Sigma,\G)$ . Setting the moment map to the zero value (flat connections!) and factoring the corresponding $0$-level 
set by the partial gauge group $G_\partial(\Sigma)$ we obtain the principal actor of our game:
\be M(\Sigma)\equiv   \Omega^1(\Sigma,\G)// G_\partial(\Sigma)\ee
It was proved in \cite{Don}, that in the case of non-empty boundary the moduli space $M(\Sigma)$ is a smooth symplectic manifold. Needless to say, for $\Sigma$ being the annulus, $M(\Sigma)$ is the phase space of the standard WZNW model.

\medskip

\noi Denote by  $G(\partial\Sigma)$ the factor group $G(\Sigma)/G_\partial(\Sigma)$.
Obviously,  $G(\partial \Sigma)$  can be identified with the group of smooth  maps from
 $\partial \Sigma$ to $G$ and it acts on the moduli space $M(\Sigma)$ in the Hamiltonian way. The equivariant moment map of this residual action is given  by the second term on the r.h.s. of (\ref{mmp}) where $A$ is the restriction of the  representant  of the class $[A]\in M(\Sigma)$ to  $\Omega^1(\partial\Sigma,\G)$.   
 
 \medskip
 
 \noi If the boundary $\partial\Sigma$ has $r+1$ connected components then to each one corresponds the equivariant  Hamiltonian action of a copy of the loop group $LG$  on $M(\Sigma)$.  
 The explicit description of the manifold $M(\Sigma)$ with $k$-handles was given in \cite{MW1} as
 \be M(\Sigma)=\biggl\{(a,c,\zeta)\in G^{2k}\times G^r\times (L\G^*)^{r+1}\bigg\vert \prod_{i=1}^{2k}[a_{2i-1},a_{2i}]=\prod_{i=1}\textrm{Hol}(\zeta_i)\biggr\}.\label{mw7}\ee
 where $c_0=e$ and $[.,.]$ stands for the group commutator. In this description, the action of $h=(h_0,\dots, h_r)\in (LG)^{r+1}$ is given by 
 \be h\triangleright a_i=\textrm{Ad}_{h_0(0)}a_i, \quad h\triangleright c_j=h_0(0)c_jh_j(0)^{-1}, \quad h\triangleright \zeta_j=\textrm{Ad}_{h_j}\zeta_j-\textrm{d}h_jh_j^{-1}.\label{mw8}\ee
 The equivariant moment map is the projection to the $(L\G^*)^{r+1}$-factor. 
 
 \medskip
 
 \noi The expression for the symplectic form on $M(\Sigma)$ in the parametrization (\ref{mw7})   is complicated and it was not given in \cite{MW1}. We shall find  now an alternative description of the space  $M(\Sigma)$  in which the  structure of the symplectic form becomes transparent and it is given in terms of the quasi-Hamiltonian fusion.

 \medskip
 
 \noi  {\bf Theorem 5}: {\it Let $\Sigma$ be a Riemann surface with $k$ handles and $r+1$ boundaries. Then
 \be M(\Sigma)=\biggl(\underbrace{W^-\fus W^-\fus\dots\fus W^-}_{r +1\ \ {\rm times}}\fus \underbrace{\ifd\fus\ifd\fus\dots\fus\ifd}_{k\ \ {\rm times}}\biggr)_e,\label{bb}\ee
where  $\ifd$  is the  internally fused quasi-Hamiltonian double of $G$.}
 
 \begin{proof}
 We shall start with the quasi-Hamiltonian $\underbrace{G\times G\times ... \times G}_{r+1 \ \ {\rm times}}$-equivalent 
 of the Hamiltonian $\underbrace{LG\times LG\times ... \times LG}_{r+1 \ \ {\rm times}}$-space $M(\Sigma)$  as obtained in \cite{AMM}:
 \be \textrm{Hol} (M(\Sigma))=\underbrace{D(G)\fus D(G)\fus.. \fus D(G)}_{r\ \ {\rm times}}\fus \underbrace{\ifd\fus\ifd\fus\dots\fus\ifd}_{k\ \ {\rm times}}.\label{pod}\ee 
 Here $D(G)$ is the standard quasi-Hamiltonian double. Using Theorem 4 and Remark 2, the  quasi-Hamiltonian representation (\ref{bb}) of $M(\Sigma)$ follows directly.
 \end{proof}
 
 \noi {\bf Corollary}: {\it The moduli space of flat connections on the surface with  $n$  boundaries, $m$ Wilson lines insertions and $k$ handles reads:
\be M_{nmk}(\Sigma)\equiv  (\underbrace{W^-\fus\dots \fus W^-}_{n \ \ {\rm times}}  \fus \ C^-_1\fus C^-_2\fus \dots \fus C^-_m\fus \underbrace{ \bod(G)\fus\dots\fus \bod(G)}_{k \ \ {\rm times}})_e.\label{BB}\ee}
\begin{proof}  Suppose that $r+1=n+m$. First of all, we  convert  into $W^-$ via Theorem 4 and Remark 2 only $n-1$ factors $D(G)$ in the fusion product  (\ref{pod}). Then  we use the fact proved in \cite{AMM},  that the inclusion of the Wilson line with holonomy in a conjugacy class $\C_i$ amounts to the reduction of $D(G)$  at  $\C_i$  and it is equal to $\C_i^-$.  On the remaining $m$ factors $D(G)$
 we thus perform the reduction at a tuple of the conjugacy classes $\C=(\C_1,...,\C_m)$  to obtain the desired formula (\ref{BB}).
\end{proof}

 \noi  
\medskip
 \section{Symplectic geometry of defects}
 
 So far we have been dealing with the side $AC$ of the triangle on Figure 1 and we have proved  the     quasi-Hamiltonian formula (\ref{B})   expressing the symplectic structure of the
 moduli space of flat connections on the surface with  $n$  boundaries, $m$ Wilson lines insertions and $k$ handles. We shall now turn to the side $BC$ of the triangle and perform the explicit evaluation of the symplectic structures of several particular WZNW defects starting from  the formula (\ref{B}) and applying successively    the formula
(\ref{mix}).  In all cases, we shall find the perfect   agreement with the results   obtained before from the detailed analysis of the WZNW dynamics  \cite{G91,G01,GTT,S}. This fact  confirms  that the concept of the quasi-Hamiltonian fusion is the unique structural
ingredient explaining the multitude of terms in the defect symplectic forms.  Before doing the actual calculations, we  should comment on two things: 
  1) We note  that the fusion product  introduced in Section 2 is commutative only on the {\it isomorphism classes}
 of quasi-Hamiltonian spaces.   Although the isomorphism between  $M_1\fus M_2$ and $M_2\fus M_1$ 
is described explicitely in \cite{AMM},  in practice it  turns out to be more convenient to reshuffle the order of the fused manifolds 
 to ensure a direct comparison of the symplectic forms issued 
from the formula (\ref{B})  with the symplectic forms of the corresponding WZNW defects  as obtained previously in  \cite{G91,G01,GTT,S}. 2) In physical literature   the loop group actions on symplectic manifolds  related to the
WZNW dynamics are often considered with the anti-equivariant moment map  \cite{G91,G01,GTT,S}. We already know from Section 4, that this is a mere convention since  the loop reversal map changes the anti-equivariant  moment
map into the equivariant.  In order to match  the same convention, sometimes we perform the  transition  from equivariant to anti-equivariant at the level of the formula (\ref{B}). As it was proved in Theorem 4, this 
 amounts  simply to replacing $W^-$ by $W$.

\medskip

\begin{enumerate}

 \item  {\it Bulk  WZNW model with no defects. }

\medskip

\noi This is an important  warm up case to start with. It is well-known that the phase space of the standard
WZNW model is the moduli space of flat connections on the surface with two boundaries  \cite{EMSS}.  Conventionally, the action of the loop group corresponding to one of the boundaries is taken to be anti-equivariant and the other one equivariant.  Following our main formula (\ref{B}),
this phase space should  therefore coincide with the symplectic manifold $(W\fus W^-)_e$. Let us see that this is indeed true.  Following the fusion formula (\ref{mix})
we obtain
\be \Omega_{W\fus W^-}=\Omega(l)-\Omega(r)- \jp(\mu_l^{-1}\delta \mu_l ,\mu_r^{-1} \delta\mu_r).\label{mav}\ee
Note the opposite sign of $\Omega(r)$ and the inverse of the moment map $\mu_r$ related to the fact that the right sector correspond to the inverse quasi-Hamiltonian
space $W^-$ in the sense explained in Section 2.   Now the quasi-Hamiltonian reduction of  $W\fus W^-$ at the unit level of the fused moment map $\mu_l\mu_r^{-1}=e$ 
makes the last term on the r.h.s. of (\ref{mav}) disappear and we are left with
\be  \Omega_{W\fus W^-}\bigg\vert_{\mu_l=\mu_r}=\Omega(l)-\Omega(r),\ee
where the configurations $l,r\in W$ have the same monodromies. But this coincides with the  expression of the standard WZNW symplectic form $\omega_{WZ}$  in terms
of the left and right movers as given by (\ref{gaw}) (cf. also \cite{G91}).

\medskip

\item{\it Bulk WZNW model  with one defect.}

\medskip

\noi The defect in the bulk WZNW model means that the WZNW configuration field $g(\sigma)$ is allowed to jump at some point $\sigma_0$ of the loop; we choose   $\sigma_0=0$.
As shows the analysis of \cite{FSW}, the preservation of the full $LG\times LG$ symmetry  requires that  the jump of the configuration field    must
lie in some conjugacy class $\C\subset G$.  Following our general philosophy of relying on  the  formula (\ref{B}),   the WZNW symplectic form in the
presence of the defect should be given by the  unit-level reduction of the fusion product $W\fus W^-\fus \C$.   Following the formulae (\ref{B}), (\ref{fcc}),   (\ref{mix})  and  
 (\ref{foo}), we find that  the quasi-Hamiltonian form on $W\fus W^-\fus \C$  restricted to the unit value of the product moment 
map $\mu_l\mu_r^{-1}\mu$ reads: \be \Omega_{W\fus W^-\fus \C}\bigg\vert_{\mu=\mu_r\mu_l^{-1}}=\Omega(l)-\Omega(r)  +\alpha^{\C}_{\mu_r\mu_l^{-1}}
 +\jp(\mu_l^{-1}\delta \mu_l ,\delta\mu_r ^{-1}\mu_r).\label{nif}\ee
Here we recall that  $\alpha^{\C}_{\mu_r\mu_l^{-1}}$ is the quasi-Hamiltonian form  (\ref{fcc}),(\ref{alt}). Our expression (\ref{nif}) coincides with Eq. (106) of \cite{S}, where the symplectic structure of the  WZNW model with one defect  was described ($\mu_{l,r}\equiv \gamma_{L,R}^{-1}$, $\alpha^{\C}=-\jp\omega$ in the notation of \cite{S}).

\item{\it Bulk WZNW model with two defects.}
 
\medskip

\noi Again from the formulae (\ref{B}), \ref{fcc}),  (\ref{mix}) and  
  (\ref{foo}), we find the quasi-Hamiltonian form on the fusion product $W\fus W^-\fus \C_1\fus \C_2$ restricted to the unit value of the product moment 
map $\mu_l\mu_r^{-1}\mu_1\mu_2$:
\be \Omega_{W\fus W^-\fus \C_1\fus \C_2}\bigg\vert_{\mu_r\mu_l^{-1}=\mu_1\mu_2}=\Omega(l)-\Omega(r)   +\alpha_{\mu_1}^{\C_1}  +\alpha_{\mu_2}^{\C_2} 
 +\jp(\mu_l^{-1}\delta \mu_l ,\delta\mu_r ^{-1}\mu_r)  +\jp( \mu_1^{-1}\delta\mu_1,\delta\mu_2 \mu_2^{-1}).\label{seq}\ee
This expression is equivalent to  Eq. (121) of \cite{S}, where the symplectic structure of the  WZNW model with two defects  was derived. To see this, some more work is needed. First of all, we have to identify the notations here and in \cite{S}:
$\mu_{l,r}\equiv \gamma_{L,R}^{-1}$, $\mu_1\equiv \tilde d_\beta$ and  $\mu_{2}\equiv d_\alpha$. Then we have to reexpress $\Omega(l)$ 
in terms of  the parametrization (\ref{bla})  of $l\in W$: 
\be l(\sigma)\equiv h(\sigma)e^{\ri \tau \sigma} g_0^{-1},\label{parr}\ee
where $h(\sigma)$ is strictly periodic (therefore it is an element of $LG$), $\tau$ is in the Weyl alcove of $\G$ and $g_0$ is in $G$.  With this parametrization, we obtain
$$ \Omega(l)  = \frac{1}{ 2}\int_0^{2\pi}\biggl[(h^{-1}\delta h, \partial_\sigma(h^{-1}\delta h))-2\ri \delta(\tau,h^{-1}\delta h)\biggr]d\sigma +$$\be + 2\pi\ri(\delta \tau, g_0^{-1}\delta g_0)+\jp(g_0^{-1}\delta g_0,e^{2\pi\ri\tau} g_0^{-1}\delta g_0 e^{-2\pi\ri\tau}). \label{aux}\ee
Inserting (\ref{aux}) into (\ref{seq}), we obtain the formula which coincides with Eq. (121) of \cite{S}.

\item  {\it Boundary WZNW model with open string ending on the conjugacy classes $\C_1$ and $\C_2$. }

\medskip

It was found in \cite{GTT}, that the symplectic structure of the boundary WZNW model with open string ending on two conjugacy classes $\C_1$ and $\C_2$  is the same as that of the moduli
space of flat connections on the disc with two Wilson lines insertions with the holonomies in $\C_1$ and $\C_2$.  Following  our quasi-Hamiltonian dictionnary, we shall evaluate the unit-level reduction of the fusion product  $\C_2\fus \C_1^-\fus W$. Thus,
assembling  the quasi-Hamiltonian form on the conjugacy classes (\ref{fcc}),  the expression of the quasi-Hamiltonian form issued from
the fusion (\ref{mix}) and the chiral WZNW form (\ref{foo}), we find the following formula for the quasi-Hamiltonian form on the fusion product $\C_2\fus \C_1^-\fus W$ restricted to the unit value of the product moment 
map $\mu_2\mu_{1}^{-1}\mu_{l}$:
\be \Omega_{\C_2\fus \C_1^-\fus W}\bigg\vert_{\mu_l^{-1}\mu_1=\mu_2}=\Omega(l) + \jp(\mu_1 \delta \mu_1^{-1},\delta\mu_l \mu_l^{-1}) -\alpha^{\C_1}_{\mu_1}+\alpha^{\C_2}_{\mu_2}.\ee
This expression coincides with Eq. (53) of \cite{GTT}, where the symplectic structure of the boundary WZNW model was first determined (to see it, one must identify $\mu_l^{-1}\equiv \gamma$,  $\mu_1\equiv h_0$ and $\mu_2\equiv \gamma h_0$).

\item{\it Boundary WZNW model with one defect.}

\medskip

\noi From the formulae  (\ref{B}), (\ref{fcc}),  (\ref{mix}) and  
  (\ref{foo}), we find the quasi-Hamiltonian form on the fusion product  $\C_1^-\fus \C_2\fus W\fus \C_3$  restricted to the unit value of the product moment 
map $\mu_1^{-1}\mu_2\mu_l\mu_3$:
\be \Omega_{\C_1^-\fus \C_2\fus W\fus \C_3}\bigg\vert_{ \mu_1=\mu_2\mu_l\mu_3}=\Omega(l)+\frac{1}{2}(\mu_1\delta \mu_1^{-1} ,\delta\mu_2\mu_2^{-1}) +\jp (\mu_l^{-1}\delta \mu_l, \delta  \mu_3\mu_3^{-1})-\alpha^{\C_1}_{\mu_1}+\alpha^{\C_2}_{\mu_2}+\alpha^{\C_3}_{\mu_3}.\label{bdd}\ee
In order that this expression coincide  with Eq. (139) of \cite{S}, where the symplectic structure of the  boundary  WZNW model with one defect  was computed, we
must  insert 
  (\ref{aux}) into (\ref{bdd}) and identify $\mu_1\equiv h_0$, $\mu_2\equiv d_\alpha$, $\mu_3\equiv h_\pi$ and $\mu_l\equiv\gamma^{-1}$.

\end{enumerate}

\end{document}